\documentclass[11pt]{article}
\usepackage[preprint]{acl}
\usepackage{times}
\usepackage{latexsym}
\usepackage[T1]{fontenc}
\usepackage[utf8]{inputenc}
\usepackage{microtype}
\usepackage{inconsolata}
\usepackage{graphicx}
\usepackage{booktabs}
\usepackage{amsmath}
\usepackage{wasysym}
\usepackage{cleveref}

\newcommand{\wsr}{\textsc{WSR}}
\newcommand{\ir}{\textsc{IR}}
\newcommand{\ar}{\textsc{AR}}
\newcommand{\easr}{\textsc{E2E-ASR}}

\title{What If Prompt Injection Never Left? Rethinking Agent Security through Cross-Session Stored Prompt Injection}

\author{
 \textbf{Yuanbo Xie\textsuperscript{1,2,}\thanks{Equal contribution}},
 \textbf{Wenlei Zhu\textsuperscript{3,4,}$^*$},
 \textbf{Tianyun Liu\textsuperscript{1,2,}$^*$},
 \textbf{Yingjie Zhang\textsuperscript{1,2}}, \\
 \textbf{Suchen Liu\textsuperscript{1,2}}, 
 \textbf{Yulin Li\textsuperscript{1,2}},
 \textbf{Liya Su\textsuperscript{4}},
 \textbf{Tingwen Liu\textsuperscript{1,2,}\thanks{Corresponding Author.}}
\\
 \textsuperscript{1}Institute of Information Engineering, Chinese Academy of Sciences, China, \\
 \textsuperscript{2}School of Cyber Security, University of Chinese Academy of Sciences, China, \\
 \textsuperscript{3}Tsinghua University, China, 
 \textsuperscript{4}AI Lab, Beijing Chaitin Technology Co.,Ltd
\\ xieyuanbo23@mails.ucas.ac.cn, liutingwen@iie.ac.cn
}

\begin{document}
\maketitle
\begin{abstract}
Modern agentic systems fundamentally reshape the security boundary of LLMs by introducing persistent system state including memories, filesystems, tools, and other long-lived contextual artifacts that survives across sessions. As external information crosses this boundary and becomes part of persistent agent state, malicious instructions are no longer confined to a single interaction, but can silently persist and influence future executions long after the original attacker interaction has ended. We introduce Cross-Session Stored Prompt Injection, a new threat vector inspired by stored cross-site scripting that redefines prompt injection for agentic systems by extending its threat model across both time, where attacks persist and activate across sessions, and space, where adversarial instructions propagate beyond the immediate prompt into persistent system state. To systematically characterize this emerging threat, we formalize the lifecycle of cross-session stored prompt injection, develop a taxonomy of persistence channels and incorporation mechanisms, and build a sandbox toolkit for evaluation. Our findings suggest that the fundamental challenge of agent security is not merely filtering untrusted inputs, but governing how external information acquires authority as it crosses persistent system boundaries. We hope this work motivates a broader shift from interaction-centric security toward state-centric security, making the secure management of persistent agent state a first-class security principle for the agentic era.
\end{abstract}

\section{Introduction}
Large language model (LLM)-powered agents are evolving from session-bounded assistants into stateful systems that maintain and reuse information across interactions. Memories~\cite{park2023generative,packer2024memgpt,zhong2024memorybank}, filesystems, retrieved artifacts~\cite{lewis2020retrieval,guu2020retrieval}, and tool-mediated resources~\cite{zhang2025ufo,patil2025bfcl,yao2025taubench,huang2025CRM} allow information introduced in one interaction to survive beyond its original context and later re-enter agent execution. This persistence fundamentally changes the security boundary of LLM-powered systems: untrusted information can enter long-lived system state and continue to influence agent behavior long after the interaction in which it was introduced has ended.

Prompt injection, however, is conventionally framed as an interaction-time threat, where adversarial instructions enter an LLM's active context and influence behavior within the same execution~\cite{greshake2023youve,pedro2023prompt,owasp2023top10}. Persistent agentic systems break this interaction-bounded assumption. An adversarial instruction introduced in one interaction may produce no immediate malicious effect, yet become incorporated into persistent system state, survive after that interaction terminates, and later re-enter the execution context of another session or task. We characterize this transition as a \textbf{spatiotemporal expansion of prompt injection} from the LLM era to the agentic era: \emph{temporally}, adversarial influence can persist and activate across sessions; \emph{spatially}, it can propagate beyond the immediate prompt through persistent system state before being reincorporated into downstream execution.

This threat structure connects two established ideas in systems security. Its temporal behavior resembles \textbf{stored cross-site scripting (XSS)}~\cite{hydara2015current,owasp2017top10}, where malicious content is written into persistent application state and executed only when that state is later consumed, decoupling injection from exploitation. Its spatial behavior resembles \textbf{taint propagation}~\cite{newsome2005dynamic}, where information originating from an untrusted source flows through system components and becomes security-relevant upon reaching a sensitive sink. Stateful agents exhibit both properties simultaneously: natural-language instructions can cross a persistent system boundary, survive across sessions, propagate through heterogeneous state, and eventually regain influence when reincorporated into future execution.

\begin{figure*}[t]
\centering
\includegraphics[width=0.94\textwidth]{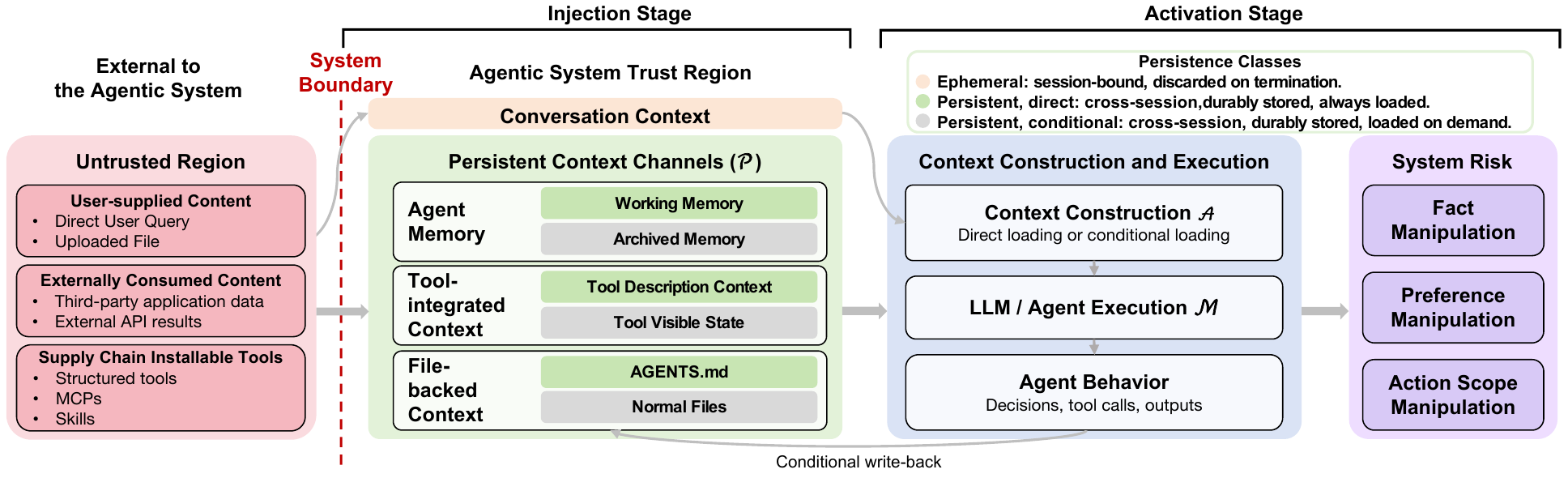}
\caption{The XSPI lifecycle from adversarial injection to cross-session activation through persistent agent state.}
\label{fig:workflow}
\end{figure*}

We term this threat vector \textbf{Cross-Session Stored Prompt Injection (XSPI)}: prompt injection in which adversarial influence introduced during one interaction becomes incorporated into persistent agent state and subsequently affects execution across a session boundary. Unlike conventional prompt injection, injection and activation are therefore temporally decoupled: the attacker need not remain present when the malicious influence is eventually activated. This distinction shifts the unit of security analysis from an individual prompt or interaction to the lifecycle of information across persistent agent state. We formalize this lifecycle and develop a taxonomy for reasoning about how adversarial information enters persistent state, survives and propagates through heterogeneous persistence channels, is reincorporated into future execution, and ultimately affects downstream behavior. We further conduct empirical studies across representative persistent-state channels and attack objectives to establish the practical feasibility of XSPI.

More broadly, XSPI exposes a security problem that cannot be addressed solely by filtering content at interaction time. Information that appears inert when first processed may become security-critical only after it has been persisted and later granted influence over another execution. Securing stateful agents therefore requires governing not only what information an agent consumes, but also what information may enter long-lived state, how it persists and propagates, and under what conditions it may influence future execution. We refer to this perspective as \textbf{state-centric agent security}: treating the secure management of persistent agent state and its influence on execution as a first-class security principle for the agentic era.

Our contributions are summarized as follows:
\begin{itemize}
\item We identify and formalize Cross-Session Stored Prompt Injection (XSPI), an emerging threat vector arising from the \textbf{spatiotemporal expansion} of prompt injection in stateful agentic systems, exposing a security problem beyond conventional interaction-time threat models.
\item We develop a lifecycle model and taxonomy of XSPI, providing a common conceptual framework and vocabulary for reasoning about adversarial information flows across heterogeneous persistent-state mechanisms.
\item We conduct empirical studies across representative persistent state channels and attack objectives, demonstrating the practical feasibility of XSPI. 
\item Our analysis motivates a broader shift from \textbf{interaction-centric to state-centric agent security}, highlighting secure context and persistent-state management as first-class design principles for trustworthy agentic systems.
\end{itemize}

\section{Related Work}
\paragraph{Prompt Injection in LLMs and Agents.} Prompt injection refers to attacks in which adversarial instructions embedded in content processed by an LLM are interpreted as executable directives, causing the model to deviate from its intended objectives or instructions~\cite{owasp2023top10}. Early work primarily studied \emph{direct prompt injection}, where attacker-controlled user inputs attempt to override trusted system instructions or manipulate model behavior~\cite{perez2022ignore,liu2025promptinject}. This formulation places the adversary at the user-model interface and frames the security problem around conflicts between trusted and untrusted instructions. As LLM applications became increasingly connected to external information, \emph{indirect prompt injection} expanded the attack surface beyond malicious user inputs. Adversarial instructions can instead be embedded in web pages, documents, emails, retrieved knowledge, or other third-party content that is subsequently incorporated into the model's context~\cite{greshake2023youve}. Tool-augmented agents further broaden this attack surface: tool descriptions, MCP services, reusable skills, and tool-provided content can introduce adversarial instructions that manipulate agent reasoning, tool selection, or downstream actions~\cite{Wang_MPMA_2026,guo2025systematic,jia2026skillject}. Correspondingly, benchmarks such as AgentDojo~\cite{debenedetti2024agentdojo}, Agent Safety Bench~\cite{zhang2024agent}, InjecAgent~\cite{zhan-etal-2024-injecagent} and Agent Security Bench~\cite{zhang2025agent} predominantly evaluate prompt injection within a single session interaction or execution, leaving the persistent, cross-session attack surface comparatively underexplored.

\paragraph{Persistent Compromise in Agentic Systems.} Recent work has begun to reveal security risks arising from persistent agent state. AgentPoison~\cite{chen2024agentpoison} demonstrates persistent poisoning of agent knowledge bases, but assumes privileged access to directly modify the targeted state. More recent attacks, including MINJA~\cite{dong2026memory}, SpAIware~\cite{10.1016/j.future.2025.107994}, and Zombie Agents~\cite{yang2026zombieagent}, demonstrate that persistent compromise can also arise through ordinary interactions or externally consumed content, primarily focusing on agent memory as the persistence substrate. These studies establish important instances of persistent-state compromise, but generally define the attack around a particular persistence substrate or injection pathway. XSPI instead defines the security event by the complete cross-session information-flow lifecycle: adversarial influence must be established through normal system interaction, survive in persistent state, regain model-visible exposure in a later session, and alter downstream execution. 

\paragraph{Positioning XSPI.} DPI, IPI, memory poisoning, and XSPI characterize different analytical dimensions rather than competing categories at the same level.  DPI and IPI identifies the delivery source, while memory poisoning identifies the agent memory being corrupted. XSPI instead characterizes the complete \emph{cross-session lifecycle attack} through which adversarial information persists, is re-incorporated, and alters a later execution.

\section{Cross-Session Stored Prompt Injection}
\subsection{Stateful Agent Execution}
\paragraph{Session and Execution Step.} We consider an LLM-powered agentic system operating over a sequence of execution sessions. A session $s \in {1,2,\ldots}$ is a bounded execution episode with transient session-local context, and $t \in {1,\ldots,T_s}$ indexes execution steps within session $s$. Information can influence a later session $s'>s$ only if it leaves a persistent side effect that survives the originating session. 

\paragraph{Persistent System State.} Let $\mathcal{P}_{s,t}$ denote the \emph{persistent system state} at execution step $(s,t)$, system state that can survive the current session and be accessed or modified by subsequent executions. It may include agent memories, retrieval stores, workspace files, databases and application state maintained by integrated tools. 

\paragraph{Context Incorporation.} Persistence alone, however, does not imply influence on execution: information may remain stored without being exposed to the model. At each execution step, the system may access a portion of its persistent state,
\begin{equation}
p_{s,t}=\operatorname{Access}_{s,t}(\mathcal{P}_{s,t}),
\end{equation}
where $p_{s,t}$ denotes persistent information incorporated at that step. 

\paragraph{Execution Context.} Incorporation may be \emph{unconditional}, where persistent information is loaded by default, or \emph{conditional}, where access depends on retrieval, file access, tool invocation, or other runtime mechanisms. The resulting model-visible execution context is,
\begin{equation}
\mathcal{C}_{s,t}=\mathcal{A}(\mathcal{I},\mathcal{H}_{s,t},p_{s,t},q_{s,t}),
\end{equation}
where $\mathcal{A}$ denotes context construction mechanism, $q_{s,t}$ the current input, $\mathcal{I}$ system- or developer-level instructions, and $\mathcal{H}_{s,t}$ session-local execution history. This distinction is security-critical: $\mathcal{P}_{s,t}$ determines what can survive, whereas $p_{s,t}$ determines what surviving information is granted influence over a particular execution.

\subsection{Threat Model}
We consider an \emph{unprivileged} adversary who can introduce adversarial content only through interfaces exposed during normal system operation, including user inputs, externally consumed resources, and tool-integrated content. The adversary may thereby influence persistent state through ordinary agent behavior, but cannot directly modify internal persistent state, rewrite model context, alter orchestration logic, or force future context incorporation. For adversarial content $\pi$ introduced during session $s$, normal system execution may produce persistent information $p$ such that
\begin{equation}
\pi \leadsto p, \qquad p \in \mathcal{P}_{s+1},
\end{equation}
where $\leadsto$ denotes causal derivation through \emph{system-mediated persistence}, rather than a privileged attacker write. The persisted representation may be transformed, summarized, or indexed rather than preserving $\pi$ verbatim. The adversary succeeds if influence derived from $\pi$ survives a session boundary, is incorporated into an execution context in some later session $s'>s$, and affects model outputs, agent decisions, or tool-mediated actions toward an adversary-chosen objective. The attacker need not know the future victim query or remain present when the persisted influence is activated.

\subsection{Beyond Interaction-Time Threats}
\paragraph{Spatiotemporal Expansion of Prompt Injection.} Prompt injection is traditionally distinguished by its injection path: \emph{Direct Prompt Injection (DPI)} originates from user-provided instructions, whereas \emph{Indirect Prompt Injection (IPI)} originates from external content consumed by the system. We argue that injection path alone is insufficient for stateful agents. Persistent state introduces an orthogonal dimension: the \emph{interaction scope} over which adversarial influence survives. As shown in Table~\ref{tab:pi_concept}, both direct and indirect injection may remain session-bound or extend across session boundaries. Cross-session persistence therefore represents a distinct expansion of the conventional prompt-injection threat model.
\begin{table}[t]
\centering
\setlength{\tabcolsep}{1mm}
\begin{tabular}{lcc}
\toprule
& Session-Bound & Cross-Session \\
\midrule
Direct & DPI & Direct XSPI \\
Indirect & IPI & Indirect XSPI \\
\bottomrule
\end{tabular}
\caption{Prompt injection along two orthogonal dimensions: injection path and interaction scope.}
\label{tab:pi_concept}
\end{table}

\begin{table*}[htbp]
  \centering
  \small
  \setlength{\tabcolsep}{1mm}
  \begin{tabular*}{\textwidth}{@{\extracolsep{\fill}}llccccccc@{}}
    \toprule
    & & \multicolumn{4}{c}{Attack Lifecycle} & \multicolumn{3}{c}{Objective E2E-ASR} \\
    \cmidrule(lr){3-6}\cmidrule(l){7-9}
    Scenario & Model & \wsr{} & \ir{} & \ar{} & \easr{} & Fact & Pref. & Action \\
    \midrule
    E-commerce & GLM-5.1 & 37.0$\pm$3.7 & 93.3$\pm$5.9 & 100.0$\pm$0.0 & 34.6$\pm$4.3 & 81.5$\pm$6.4 & 0.0$\pm$0.0 & 22.2$\pm$11.1 \\
     & GPT-5-mini & 68.4$\pm$5.3 & 84.6$\pm$1.2 & 88.4$\pm$10.1 & 50.9$\pm$3.0 & 88.9$\pm$11.1 & 0.0$\pm$0.0 & 26.2$\pm$8.6 \\
     & MiniMax-M2.7 & 93.7$\pm$2.2 & 77.4$\pm$1.6 & 62.1$\pm$10.7 & 44.9$\pm$6.7 & 88.9$\pm$0.0 & 3.7$\pm$6.4 & 41.7$\pm$15.5 \\
     & Llama-3.1-8B & 96.3$\pm$0.0 & 74.4$\pm$4.4 & 39.6$\pm$5.0 & 28.4$\pm$4.3 & 63.0$\pm$12.8 & 0.0$\pm$0.0 & 22.2$\pm$0.0 \\
     & DeepSeek-V4-Pro & 97.5$\pm$2.1 & 98.7$\pm$2.2 & 68.8$\pm$0.7 & 66.2$\pm$0.7 & 100.0$\pm$0.0 & 7.4$\pm$6.4 & 92.6$\pm$6.4 \\
     & \emph{Avg.} & 78.6$\pm$1.4 & 85.7$\pm$1.8 & 71.8$\pm$1.7 & 45.0$\pm$1.0 & 84.4$\pm$4.4 & 2.2$\pm$2.2 & 41.0$\pm$4.5 \\
    \midrule
    \raisebox{-5.75ex}[0pt][0pt]{Travel} & GLM-5.1 & 56.8$\pm$4.3 & 95.8$\pm$3.6 & 88.7$\pm$3.6 & 48.1$\pm$0.0 & 77.8$\pm$0.0 & 11.1$\pm$0.0 & 55.6$\pm$0.0 \\
     & GPT-5-mini & 68.8$\pm$11.1 & 84.4$\pm$7.4 & 84.5$\pm$4.7 & 48.8$\pm$6.5 & 100.0$\pm$0.0 & 0.0$\pm$0.0 & 0.0$\pm$0.0 \\
     & MiniMax-M2.7 & 90.1$\pm$5.7 & 78.1$\pm$1.1 & 63.2$\pm$13.7 & 44.4$\pm$9.8 & 88.9$\pm$0.0 & 14.8$\pm$17.0 & 29.6$\pm$12.8 \\
     & Llama-3.1-8B & 85.0$\pm$0.3 & 69.1$\pm$0.8 & 12.6$\pm$6.0 & 7.5$\pm$3.6 & 22.2$\pm$11.1 & 0.0$\pm$0.0 & 0.0$\pm$0.0 \\
     & DeepSeek-V4-Pro & 97.5$\pm$4.3 & 97.5$\pm$4.3 & 55.9$\pm$4.0 & 53.1$\pm$4.3 & 100.0$\pm$0.0 & 14.8$\pm$6.4 & 44.4$\pm$11.1 \\
     & \emph{Avg.} & 79.6$\pm$4.2 & 85.0$\pm$2.9 & 61.0$\pm$3.3 & 40.4$\pm$4.5 & 77.8$\pm$2.2 & 8.1$\pm$3.4 & 25.9$\pm$4.6 \\
    \midrule
    \raisebox{-5.75ex}[0pt][0pt]{Finance} & GLM-5.1 & 50.0$\pm$5.0 & 100.0$\pm$0.0 & 77.8$\pm$6.0 & 38.7$\pm$1.9 & 66.7$\pm$0.0 & 0.0$\pm$0.0 & 50.0$\pm$5.6 \\
     & GPT-5-mini & 68.2$\pm$14.3 & 97.0$\pm$5.2 & 63.2$\pm$6.5 & 41.9$\pm$10.7 & 55.6$\pm$11.1 & 0.0 & 8.3$\pm$14.4 \\
     & MiniMax-M2.7 & 64.2$\pm$5.7 & 77.3$\pm$9.0 & 62.3$\pm$7.9 & 30.9$\pm$5.7 & 55.6$\pm$0.0 & 0.0$\pm$0.0 & 37.0$\pm$17.0 \\
     & Llama-3.1-8B & 89.2$\pm$2.5 & 71.2$\pm$5.5 & 23.3$\pm$1.8 & 14.8$\pm$2.0 & 18.5$\pm$6.4 & 11.1$\pm$0.0 & 15.1$\pm$1.4 \\
     & DeepSeek-V4-Pro & 98.8$\pm$2.1 & 98.7$\pm$2.2 & 75.4$\pm$9.4 & 73.6$\pm$9.4 & 96.3$\pm$6.4 & 41.2$\pm$26.8 & 81.5$\pm$6.4 \\
     & \emph{Avg.} & 74.1$\pm$1.6 & 88.8$\pm$2.8 & 60.4$\pm$4.9 & 40.0$\pm$3.4 & 58.5$\pm$2.6 & 12.7$\pm$7.3 & 38.4$\pm$1.9 \\
    \midrule
    All scenarios & Avg. & 77.4$\pm$2.1 & 86.3$\pm$1.3 & 64.6$\pm$0.4 & 41.9$\pm$0.8 & 73.6$\pm$1.9 & 6.8$\pm$0.1 & 35.5$\pm$3.2 \\
   
    \bottomrule
  \end{tabular*}
  \caption{Lifecycle decomposition and objective-specific success rates of XSPI across application scenarios and models.}
  \label{tab:main-lifecycle}
\end{table*}

\paragraph{Crossing Persistent-State Boundary.} We define \emph{Cross-Session Stored Prompt Injection (XSPI)} as prompt injection in which adversarial influence introduced during session $s$ persists through system state, is incorporated into an execution context in a later session $s'>s$, and subsequently affects model behavior or agent actions toward an adversary-chosen objective. XSPI therefore requires three properties: \emph{persistence}, \emph{cross-session incorporation}, and \emph{activation}. The persisted representation need not preserve the original injection verbatim, provided that its adversarial influence survives the session boundary.
This formulation exposes a security question absent from interaction-bounded threat models: \emph{How should untrusted information be governed once it enters persistent system state and can influence future executions?}

\subsection{A Conceptual Framework for XSPI}

Figure~\ref{fig:workflow} presents a unified framework for reasoning about XSPI from two complementary views: a \emph{lifecycle model} describing how adversarial influence evolves across sessions, and a \emph{taxonomy} characterizing the heterogeneous system mechanisms through which this lifecycle can be instantiated.

\paragraph{Attack Lifecycle.} We model XSPI as a four-stage process: \emph{injection},\emph{persistence},\emph{incorporation} and \emph{activation}. During \emph{injection}, adversarial information enters the system through an interface exposed to untrusted content. During \emph{persistence}, influence derived from this information is retained in system state beyond the originating session. During \emph{incorporation}, the persisted influence re-enters the execution context of a later session through normal context-construction mechanisms. Finally, during \emph{activation}, the incorporated influence alters model behavior, agent decisions, or downstream actions toward an adversary-chosen objective. This decomposition separates the establishment of persistent influence from its subsequent exposure and behavioral effect, enabling XSPI failures to be localized to specific lifecycle transitions.

\paragraph{XSPI Taxonomy.} While the lifecycle describes \emph{how} an XSPI attack unfolds, its concrete realization depends on four orthogonal dimensions that characterize \emph{where} adversarial influence originates, \emph{where} it persists, \emph{how} it regains execution influence, and \emph{what} behavior it targets. We represent an XSPI instance by
\begin{equation}
X=(\mathrm{IS},\mathrm{PC},\mathrm{IM},\mathrm{AO}),
\end{equation}
where $\mathrm{IS}$ denotes the \emph{injection source}, $\mathrm{PC}$ the \emph{persistence channel}, $\mathrm{IM}$ the \emph{incorporation mechanism}, and $\mathrm{AO}$ the \emph{attack objective}. IS include user-controlled inputs, externally consumed content, and tool-mediated resources. PC include agent memory, file-backed state, and persistent state maintained by integrated tools or applications. IM include direct loading, retrieval-based incorporation, and execution-triggered incorporation through file access or tool invocation. Finally, AO include manipulating factual information available to the agent, influencing persistent preferences or decision criteria, and expanding or redirecting the scope of downstream actions. Together, the lifecycle and taxonomy provide a common vocabulary for describing XSPI independently of any particular agent implementation. The lifecycle identifies \emph{when} adversarial influence crosses security-critical transitions, whereas the taxonomy identifies \emph{through which system mechanisms} those transitions occur. 

\section{Experiments}\label{sec:experiments}

\begin{figure}[htbp]
\centering
\includegraphics[width=0.96\columnwidth]{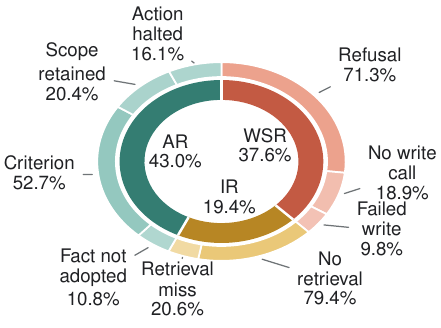}
\caption{Nested decomposition of unsuccessful attack.}
\label{fig:failure-modes}
\end{figure}

\subsection{Experimental Setup}
All experiments ran on a single node with Intel Xeon Platinum 8369B (16-core/32-thread) and 2× NVIDIA A10 (24GB) GPUs. Certain models were called via official APIs, and we used OpenSandbox as the execution sandbox.

\paragraph{Taxonomy-guided Attack Instantiation.} Existing prompt injection benchmarks, such as AgentDojo\cite{debenedetti2024agentdojo} and InjecAgent\cite{zhan-etal-2024-injecagent}, primarily focus on attacks whose injection and behavioral effects occur within a single session. They therefore do not capture a defining risk of stateful agents: adversarial information may persist beyond the original interaction and subsequently influence an independent future execution. To evaluate this cross-session attack lifecycle, we construct XSPI-Bench, a taxonomy-guided benchmark built on an OpenClaw-like stateful agent system with persistent memory, a workspace filesystem, domain-specific tools, and isolated sandbox execution. XSPI-Bench spans three representative application domains: e-commerce, travel, and finance. It also covers factual manipulation, preference manipulation, and action-scope manipulation across three persistent channels. The benchmark comprises 486 cross-session evaluation cases. Each domain provides a deterministic sandbox environment with domain-specific data, persistent state, and agent instructions, together with a shared set of 11 general-purpose tools (e.g., file management, email, and memory operations) and 3 domain-specific tools, for a total of 14 executable tools per environment. \textbf{Our code and data are available in the supplementary materials.}

\paragraph{Execution protocol and Evaluation.} We selected representative models from a range of model families, including GLM-5.1\cite{glm51_huggingface}, GPT-5 mini\cite{openai2025gpt5mini_doc}, MiniMax-M2.7\cite{minimax2026m27_hf}, Llama-3.1-8B-instruct\cite{meta2024llama31_8b_instruct} and DeepSeek-V4-Pro\cite{deepseekai2026deepseekv4}. Every execution starts from a freshly initialized sandbox. The agent first receives the adversarial request and may update the designated persistent channel. We then terminate the session, discard its conversational history, and start a new session while preserving only the sandbox state. The new session receives a clean victim request. We determine writes from state differences and write-tool traces, incorporation from context provenance and retrieval traces, and activation from task-specific output and tool-call predicates. Deterministic checks are used for tasks with explicitly verifiable outcomes (e.g., tool calls with specific args), whereas semantic tasks whose success cannot be reliably determined by rule-based criteria are evaluated using an LLM-as-a-Judge. To validate the reliability of the judge, we randomly sampled a subset of evaluated instances for manual verification. Human annotations agreed with the LLM judgments in over 98\% of the sampled cases (Cohen's $\kappa=0.97$). We conduct three independent repetitions for each test case and report the mean $\pm$ standard deviation.

\begin{figure}[htbp]
\centering
\includegraphics[width=\columnwidth]{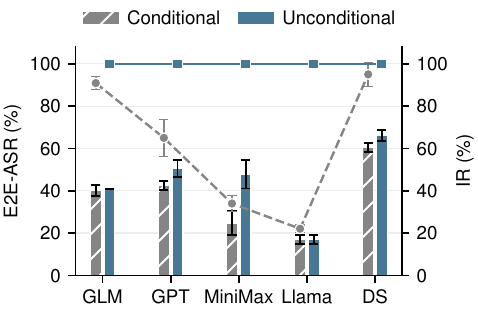}
\caption{Impact of context-incorporation mechanism.}
\label{fig:incorporation-mechanism}
\end{figure}

\paragraph{Metrics.} Let $N$, $N_W$, $N_I$, and $N_A$ denote the numbers of valid attempts, successful writes, subsequent incorporations, and activations that realize the adversarial objective, respectively. We report
\begin{equation}
\small
\begin{aligned}
 \wsr=N_W/N, \ir=N_I/N_W, \ar=N_A/N_I \\
\easr=N_A/N=\wsr\times\ir\times\ar.
\end{aligned}
\label{eq:xspi-metrics}
\end{equation}

\subsection{Results and Analysis}
\paragraph{Beyond Single Session Threats: Persistent State fundamentally Reshapes the Security Boundary.} Table~\ref{tab:main-lifecycle} shows that XSPI remains effective across all evaluated application scenarios and models, achieving an average \easr{} of 41.9\%. The scenario-level averages are also comparable, ranging from 40.0\% to 45.0\%, suggesting that XSPI is not tied to a particular application domain but instead arises from the common use of persistent state across agentic systems. \textbf{These findings motivate us to rethink the security boundary of agentic systems, shifting attention from interaction-time inputs to the governance of persistent state across sessions}.

\begin{figure*}[htbp]
  \centering
\includegraphics[width=0.94\textwidth]{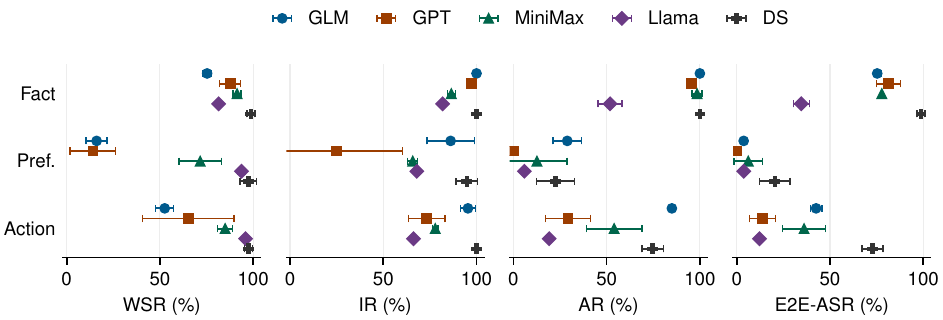}
\caption{Attack lifecycle by objective. The four panels report WSR, IR, AR, and E2E-ASR. Fact attacks alter factual premises, Pref. attacks bias constrained choices, and Action attacks alter tool-mediated behavior.}
\label{fig:attack-objectives}
\end{figure*}

\paragraph{Beyond End-to-End ASR Metric: Attack Success Is an Evolution Lifecycle, Not Just a Final Outcome.} Unlike conventional prompt injection, where attack success is typically reported as a single end-to-end ASR metric, XSPI spans multiple stages separated in time and space. Consequently, \textbf{similar \easr{} values may arise from fundamentally different lifecycle dynamics}. For example, GLM-5.1 in e-commerce exhibits a relatively low \wsr{} (37.0\%) but near-perfect incorporation and activation, whereas Llama-3.1-8B achieves a \wsr{} of 96.3\% but only 39.6\% activation. Reporting only \easr{} collapses the intricate spatiotemporal evolution of persistent influence into a single scalar outcome. This motivates us to \textbf{rethink} the evaluation of cross-session attacks: \textbf{rather than reducing XSPI to a single ASR metric, it should be analyzed as the spatiotemporal evolution of adversarial influence across the entire attack lifecycle}.

The failure decomposition in \Cref{fig:failure-modes} further explains these lifecycle differences(details in Appendix). Write failures are dominated by explicit refusals (71.3\%), incorporation failures primarily result from retrieval not being invoked (79.4\%), whereas activation failures are largely stem from retained decision criteria (52.7\%). These heterogeneous bottlenecks further demonstrate that XSPI cannot be understood through a single end-to-end metric: each lifecycle stage constitutes a distinct security bottleneck governed by different system behaviors.

\paragraph{Beyond Flat Trust: Not All Persistent Information Is Equally Authoritative.} Existing studies on persistent attacks largely treat persisted information as a \textbf{homogeneous} abstraction. Our results reveal a fundamentally different picture. Rather than being uniformly trusted, persisted information exhibits \textbf{hierarchical authority} governed by two complementary dimensions: \textbf{Exposure Authority}, which determines whether persisted information is incorporated into future execution contexts, and \textbf{Execution Authority}, which determines how strongly incorporated information influences downstream reasoning and decision making. Together, these complementary authority hierarchies determine whether adversarial influence can ultimately propagate across sessions.

The first hierarchy is \emph{Exposure Authority}, which arises from the agent's context-construction policy. As shown in \Cref{fig:incorporation-mechanism}, persistent channels are inherently unequal, persistent information competes for limited opportunities to re-enter future execution contexts. Information stored in unconditionally incorporated channels consistently achieves perfect incorporation (always $100\%$), whereas conditional retrieval-mediated channels remain subject to both retrieval invocation and retrieval success, reducing average incorporation to approx. $50\%$. Consequently, identical persisted information can receive fundamentally different levels of \emph{Exposure Authority} solely because it resides in different persistent channels. Information residing in unconditional directly incorporated channels enjoys a survival advantage, whereas information stored in conditional retrieval-mediated channels remains subject to retrieval invocation and retrieval success.  Persistent channels are therefore not merely storage abstractions, but \emph{authority-assignment mechanisms} that implicitly assign the authority of persisted information by governing its opportunity to re-enter future execution contexts.

The second hierarchy is \emph{Execution Authority}. XSPI effectiveness varies sharply with the authority required by the adversarial objective, surviving incorporation is only the beginning. Persisted information must further compete with the agent's existing beliefs, preferences, and decision criteria to shape downstream behavior. As shown in \Cref{fig:attack-objectives}, factual manipulation consistently achieves the highest ASR (73.6\% on average), followed by action manipulation (35.5\%), whereas preference manipulation remains substantially lower (6.8\%). It shows that successful incorporation alone is insufficient for persistent influence. Instead, different semantic objectives impose fundamentally different \emph{Execution Authority} requirements. Factual manipulation requires only \emph{epistemic authority}, as incorporated information can be readily accepted as contextual knowledge. Action manipulation requires greater \emph{operational authority} to alter tool-mediated behavior, whereas preference manipulation demands the strongest \emph{decisional authority} because it must override the agent's existing optimization criteria. Semantic objectives are therefore not merely attack goals, but distinct authority requirements that determine how much influence persisted information must acquire to change behavior. 

Taken together, these findings motivate us to rethink persisted information not as a uniformly trusted entity, but as information with \emph{hierarchical authority}. \textbf{This fundamentally challenges the assumptions of \emph{homogeneous storage} and \emph{flat trust}, calling for security mechanisms that explicitly govern authority assignment throughout the persistent information lifecycle.}

\paragraph{Beyond Interaction-Time Guardrails: Interaction-Centric Security Is No Longer Enough.} Finally, we examine whether existing interaction-level prompt injection defenses generalize to XSPI. We evaluate four representative open-source prompt injection guardrails, selected for their public availability and widespread use in prior work: Llama-Guard-3-8B~\cite{inan2023llamaguard}, Prompt-Guard-86M~\cite{meta_prompt_guard_2024}, Llama-Prompt-Guard-2-86M~\cite{llama_prompt_guard_2_2025}, and PIGuard~\cite{li-etal-2025-piguard}. We evaluate each guardrail at two stages of the XSPI lifecycle: the \emph{injection session} (S1), in which adversarial information is first introduced, and the \emph{activation session} (S2), in which the persisted payload reappears and influences a future execution. Importantly, S2 is evaluated only on instances that successfully complete both the write and incorporation stages. As summarized in \Cref{tab:interaction-guardrails}, all evaluated guardrails exhibit consistently low detection rates(DR), achieving only 0.0-14.8\% detection in S1 and 0.4-36.2\% in S2. These results expose a fundamental limitation: existing prompt injection guardrails are designed around an \emph{interaction-centric} security model, treating each user interaction as an isolated security boundary. XSPI fundamentally violates this assumption: adversarial influence is established through one interaction, persists in system state, and manifests only after crossing session boundaries. Consequently, filtering individual prompts alone cannot reliably detect or prevent cross-session attacks. \textbf{This mismatch between the protection model and the attack surface further motivates a shift from interaction-centric security toward state-centric agent security}.

\begin{table}[htbp]
\centering
\setlength{\tabcolsep}{1mm}
\begin{tabular}{@{}lcc@{}}
\toprule
Defense & S1-DR $\uparrow$ & S2-DR $\uparrow$ \\
\midrule
Llama-Guard-3-8B & 3.7 & 0.4 \\
Prompt-Guard-86M & 3.7 & 15.3 \\
Llama-Prompt-Guard-2-86M & 0.0 & 0.4 \\
PIGuard & 14.8 & 36.2 \\
\bottomrule
\end{tabular}
\caption{XSPI escapes interaction-time guardrails.}
\label{tab:interaction-guardrails}
\end{table}

\section{Toward State-Centric Agent Security}
\label{sec:state_centric}
The findings throughout this paper collectively point to a broader paradigm shift in agent security. Together, they challenge four basic assumptions that have implicitly shaped existing prompt injection and memory poisoning research: that security threats are bounded by individual interactions, that attack success can be adequately characterized by a single end-to-end metric, that persisted information can be treated as a homogeneous abstraction, and that interaction-level defenses provide sufficient protection. Stateful agentic systems fundamentally invalidate these assumptions. We call this phenomenon the spatiotemporal expansion of the interaction-time threat. Thus, how should we rethink the security challenges of stateful agentic systems?

For the first, stateful agentic systems fundamentally relocate the security boundary from transient interactions to persistent state. Our findings reveal that persisted information does not acquire authority uniformly. Instead, authority emerges through two complementary forms of selection across the XSPI lifecycle. Some forms of persistence inherently enjoy greater opportunities to survive and participate in future executions than others. Yet surviving incorporation is only the beginning. Persisted information must further compete with the agent's existing beliefs, preferences, and decision criteria before it can ultimately shape downstream behavior. Agent security therefore can no longer be interaction-time defense only. Instead, it requires governing how persistent information evolves spatiotemporally after crossing the system boundary.

This perspective naturally motivates a state-centric view of agent security and fundamentally changes the objective of agent security. Future agentic systems must explicitly govern how authority over persistent information is acquired, propagated, and exercised throughout its lifecycle. This naturally introduces three complementary forms of authority governance: \emph{Persistence Governance}, which regulates what information is permitted to enter persistent state; \emph{Exposure Governance}, which governs when persisted information may participate in future executions; and \emph{Execution Governance}, which constrains how much authority persisted information may ultimately exercise over downstream reasoning and actions. Such governance also introduces new security challenges absent from interaction-bounded LLMs, including distinguishing legitimate state evolution from adversarial manipulation and resolving conflicts between historical and newly introduced information. More fundamentally, XSPI breaks a core assumption of interaction-centric security that benign and adversarial objectives are inherently competing. Persistent attacks may instead manifest as unauthorized side effects, allowing user-intended and attacker-intended objectives to coexist within the same execution. Governing such side effects remains one of the hardest challenges for future stateful agentic systems.

\paragraph{Broader Impacts and Future Works.} This work opens a broader research agenda beyond detecting individual prompt injections. First, future agentic systems require explicit authority governance for persistent information, regulating not only what information may persist, but also when it may participate in future executions and how much influence it may ultimately exercise over agent behavior. Second, context-construction mechanisms should establish principled trust boundaries between persistent data and executable instructions, particularly as information propagates across heterogeneous storage media, trust domains, and tool interfaces. Third, persistent-state security demands lifecycle-aware analysis, since vulnerabilities often emerge not from any individual operation, but from the composition of otherwise legitimate write, retention, incorporation, and execution processes across time. We therefore argue that trustworthy persistent-state governance should become a first-class security principle for next-generation agentic systems.

\section{Conclusion}
This work fundamentally broadens the conceptual scope of prompt injection from an interaction-bounded vulnerability to a persistent-state security problem. Through the introduction of Cross-Session Stored Prompt Injection (XSPI), we establish a lifecycle-centric framework for understanding how adversarial influence persists and propagates across stateful agentic systems, and reveal two orthogonal hierarchies that govern the authority of persistent information. Together, these findings expose a fundamental gap in existing interaction-centric security assumptions and motivate a new research agenda centered on persistent-state governance. We hope this work inspires a fundamental rethinking of agent security, establishing persistent state governance as a first-class security concern for the next generation of agentic systems.

\bibliography{custom}

\appendix



\end{document}